\documentclass[11pt,a4paper,reqno]{amsart}

\usepackage[english]{babel}
\usepackage{siunitx}
\usepackage{natbib}
\usepackage{upgreek}
\usepackage{tabularx}
\usepackage{graphicx} 
\usepackage[headings]{fullpage}
\usepackage{amsaddr}
\usepackage[permil]{overpic}
\usepackage{amsmath}
\usepackage{float}
\usepackage{dsfont}
\usepackage{xcolor}
\usepackage{multirow}
\usepackage[singlelinecheck=false]{caption}
\usepackage{tikz}
\usetikzlibrary{shapes, arrows, positioning,decorations.markings}
\tikzstyle{vecArrow} = [thick, decoration={markings,mark=at position
    1 with {\arrow[semithick]{open triangle 60}}},
double distance=1.4pt, shorten >= 5.5pt,
preaction = {decorate},
postaction = {draw,line width=1.4pt, white,shorten >= 4.5pt}]
\tikzstyle{innerWhite} = [semithick, white,line width=1.4pt, shorten >= 4.5pt]
\usepackage{csquotes}

\newcolumntype{Y}{>{\centering\arraybackslash}X}

\newcommand{\R}{\mathbb{R}}
\newcommand{\Z}{\mathbb{Z}}
\newcommand{\N}{\mathbb{N}}

\newcommand{\gramm}{\mathrm{g}}

\newcommand{\por}{\varepsilon}
\newcommand{\vpnm}{v_{\textup{PNM}}}
\newcommand{\rmax}{r_{\textup{max}}}

\newcommand{\rate}[1]{v^{(#1)}}

\newcommand\Tstrut{\rule{0pt}{2.6ex}}         
\newcommand\Bstrut{\rule[-0.9ex]{0pt}{0pt}}

\begin{document}
\title[]{Predicting air flow in calendered paper sheets from $\upmu$-CT data: combining physics with morphology}

\author{Phillip Gräfensteiner$^{1}$, Andoni Rodriguez$^{2}$, Peter~Leitl$^{2,3}$, Ekaterina~Baikova$^{3,4}$, 
Maximilian Fuchs$^{3,4}$,
Eduardo~Machado Charry$^{3,4}$, Ulrich~Hirn$^{4,5}$, Andr\'{e}~Hilger$^6$,  Ingo~Manke$^{6}$,  Robert~Schennach$^{3,4}$, Matthias~Neumann$^{7}$,  Volker~Schmidt$^{1}$, Karin~Zojer$^{3,4}$}

\address{$^1$Institute of Stochastics, Ulm University, Helmholtzstra\ss e~18, 89069 Ulm, Germany}
\address{$^2$bionic surface technologies GmbH, Liebenauer Hauptstraße 2-6, Graz, 8041, Austria}
\address{$^3$Institute of Solid State Physics, NAWI Graz, Graz University of Technology, Petersgasse~16/II, 8010 Graz, Austria}
\address{$^4$Christian Doppler Laboratory for Mass Transport through Paper, Graz University of Technology, Petersgasse~16/II, 8010 Graz, Austria}
\address{$^5$Institute of Bioproducts and Paper Technology, Graz University of Technology, Inffeldgasse~23, 8010 Graz, Austria}

\address{$^6$Institute of Applied Materials, Helmholtz-Zentrum Berlin f\"ur Materialien und Energie, Hahn-Meitner-Platz~1, 14109 Berlin, Germany}
\address{$^7$Institute of Statistics, Graz University of Technology, Kopernikusgasse 24/III, 8010 Graz, Austria}

\email{phillip.graefensteiner@uni-ulm.de (corresponding author)}

\begin{abstract}
Predicting the macroscopic properties of thin fiber-based porous materials from their microscopic morphology remains challenging because of the structural heterogeneity of these materials.  In this study, computational fluid dynamics simulations were performed to compute volume air flow based on tomographic image data of uncompressed and compressed paper sheets.
To reduce computational demands, a pore network model was employed, allowing volume air flow to be approximated with less computational effort.
To improve prediction accuracy, geometric descriptors of the pore space, such as porosity, surface area, median pore radius, and geodesic tortuosity, were combined with predictions of the pore network model.
This integrated approach significantly improves the predictive power of the pore network model and 
indicates which aspects of the pore space morphology are not accurately represented within the pore network model.
In particular, we illustrate that a high correlation among descriptors does not necessarily imply redundancy in a combined prediction.

\end{abstract}

\maketitle

\section*{Article highlights}
\begin{itemize}
    \item Air flow through paper sheets has been determined by computational fluid dynamics and pore network model simulations
    \item Geometric descriptors of pore space are used in regression models to improve the prediction of pore network model simulations
    \item Structure-property-relationships indicate descriptors complementary to pore network modeling for different paper grades
\end{itemize}

\section{Introduction}
Predicting the macroscopic flow of gasses through a heterogeneous porous material requires knowledge of the size, shape, and connectivity of the pores involved, regardless of whether the flow is diffusive, laminar, or turbulent. 
Although methods for transport simulations exist that use 3D image data to take the geometry of the pore space into account~\cite{torquato.2013}, it remains difficult to establish relationships between morphology and transport  for flow through thin, sheet-like structures. Prominent examples of such structures are fibrous membranes encountered in paper, gas diffusion layers, or filters. In such sheets, the pore structure often exhibits strong local heterogeneities. 
To understand the flow through the total sheet area, it is necessary to capture the local flow variations across many different realizations of the pore space. This requires an enormous amount of simulations, which, in turn,  requires transport simulation methods that are accurate and, simultaneously,  can be performed with reasonable computational effort.

The choice of the simulation method depends on whether diffusive, laminar, or turbulent transport is considered.
In the case of laminar flow, there a several possible methods that, by and large, have in common that they represent different routes to predict the flow originating from the Navier-Stokes equation, reducing the transport problem to the simpler Stokes equation~\cite{Leal_2007}.
Computational fluid dynamics (CFD) and lattice Boltzmann simulations are suitable and well-established methods to determine local flows from the actual pore space~\cite{Chung_2002, succi.1991}. 
In CFD simulations, the pore phase, as supplied by tomographic image data, is represented by a surface mesh on which the Stokes equation is solved. This simulation method allows for readily considering specific situations such as stationary or incompressible flows.
Lattice Boltzmann simulations track the motion by monitoring the velocity components along a given set of directions in a cubic grid. With increasing complexity of the pore space morphology, more velocity components and smaller time steps are required to accurately solve the Boltzmann equation that is equivalent to the stationary Navier-Stokes equation.
Each of these methods is computationally demanding~\cite{yang_intercomparison_2016}.
On the other hand, ready-trained deep learning methods do not offer a route to evade the computational costs yet, as they exclusively act on the material classes they were trained for and cannot be universally applied to all types of materials, at least not without supplying costly simulations for retraining~\cite{morgan.2020}. 
However, the modeling of pore networks is a promising and computationally cheaper alternative~\cite{blunt_pore-scale_2013}.
In partitioning the pore space into distinct components, pore network modeling (PNM) preserves the transport-relevant local connections and bottlenecks between the pores. 
Transport is considered locally by solving the transport equation through neighboring pores assuming a simplified geometry of involved pores. 
In order to estimate the total flow through the entire pore space, the local transport processes are consistently connected with each other by enforcing mass conservation.

It is highly desirable to interpret or even predict the fluxes using geometric descriptors associated with the pore space, as these can be reliably determined from tomographic image data without the need of performing physical transport simulations.
Using methods from spatial statistics and mathematical morphology, it is possible to quantitatively characterize the complex morphology of the pore space~\cite{chiu.2013, jeulin.2021, ohser.2009}. In particular, to quantify local heterogeneities in paper-based materials, methods have been developed to quantify the variance and correlation of multiple local geometric descriptors computed from tomographic image data~\cite{neumann.2024}. This methodology has recently been applied to quantify local heterogeneities in polymer-based batteries~\cite{n.2022, ademmer2023unveiling, dodell2025quantifying}.

Predicting effective properties, such as permeability, based on such sets of descriptors is fundamental in many studies. 
The selected geometric  descriptors of the pore space are usually inspired by the properties used to interpret the permeability of the material, for laminar flow in the framework of Darcy’s law~\cite{whitaker_flow_1986}. 
Prominent transport-relevant descriptors are the volume fraction of the pore space (porosity), the local thickness of the material, the internal surface area per unit volume, and the mean geodesic tortuosity to quantify the winding of transportation paths. 
However, determining how many and which descriptors are truly relevant for a given porous material is not straightforward. Particularly problematic is the commonly encountered situation that these descriptors depend on each other, \emph{i.e.}, varying the value of one descriptor immediately affects the values of all other descriptors. 
The specific consequences of correlations between descriptors depend on the material, since the actual degree of correlation between descriptors can vary strongly from material to material~\cite{neumann.2024}.

In the present study, we investigate the problem of modeling laminar flow using the example of two paper samples with different morphologies. 
These paper sheets originate from the same paper grade whose structure is considered before and after compression in thickness direction.
For both samples, the air fluxes have been determined experimentally and the microstructures have been acquired by means of $\upmu$-CT measurements~\cite{n.2021}. Local variations in the microstructures are quantified via transport-relevant descriptors of pore space that were computed based on the tomographic image data. 
Furthermore, the correlation structure between transport-relevant descriptors has already been quantified in previous works  by means of  R-vine copulas~\cite{neumann.2024, n.2021}.

For each paper sample, we relate the fluxes obtained from CFD simulations to the fluxes determined by PNM simulations and various geometric descriptors, using several power-law models.
It must be emphasized that such fits do not suggest physically motivated expressions, \emph{i.e.}, the fitted exponents of the geometric  descriptors cannot guarantee trends that reveal the underlying transport physics. However, 
by means of these models, we can study microstructure-property relationships between geometric descriptors of pore space and the volumetric flux as determined by CFD simulations. Furthermore, we can evaluate the quality of the PNM simulations and monitor which descriptors improve the prediction of effective properties and thus hint towards details in pore space morphology that are not captured in PNM simulations.
In order to improve the agreement between the results of CFD and PNM simulations, we use combinations of various geometric descriptors of pore space, such as the mean geodesic tortuosity, the surface area per unit volume or scalar quantities derived from the continuous pore size distribution.

\section{Materials, tomographic imaging  and transport regime}\label{sec:modelsystem}

First, in Section~\ref{sec:material}, we give a short overview of the  materials investigated in this study and the tomographic image data measured for these materials. In Section~\ref{sec:transport_regime}, we introduce the transport regime considered in the present study that will be solved by means of CFD and PNM simulations, as will be explained later in Section~\ref{sec:transp_methods}.

\subsection{Materials and their tomographic imaging}\label{sec:material}
\begin{figure}[h]
    \centering
    \includegraphics[width=\textwidth]{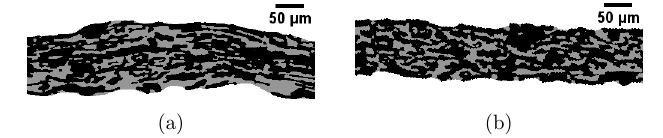}
    \caption{Slices of $\upmu$-CT data for uncompressed (a) and compressed (b) paper sheets (from cutouts \SI{500}{\micro\metre}$\times$\SI{500}{\micro\metre}$\times$height). The solid and pore phase are depicted in black and gray, respectively.}
    \label{fig:material_vis}
\end{figure}

Our study is based on two data sets of tomographic image data for two different paper sheets.
One sample type represents a paper sheet compressed in thickness direction and one represents a sheet of the paper grade before compression~\cite{n.2021}.
The latter, uncompressed sample is commercial, unbleached paper with a specific basis weight of \SI{100}{\gramm\per\metre\squared}. Compressed samples are obtained via hard-nip, steel-steel calendering of the uncompressed paper sheets with a line load of \SI{90}{\newton\per\metre}.  
The 3D microstructures of both samples are resolved by  $\upmu$-CT image data  with a voxel size of \SI{1.3}{\micro\metre} as described in~\cite{n.2021}.
For determining the boundary of the paper sheets, which particurlarly allows for computing the thickness, a rolling ball approach~\cite{sternberg.1983} is used as already performed in~\cite{n.2022, machadocharry.2018}. Exemplary slices of the segmented tomographic image data are shown in Figure~\ref{fig:material_vis}.

Since both types of (CFD and PNM) simulations performed in this study require spatially resolved 3D information of the pore phase, our common starting point is to predefine cutouts of the 3D stack of tomographic image data,
where each cutout has a cross sectional area of 
\begin{equation}\label{eq:A_seg}
    A_{\mathrm{seg}} = 500 \times \SI{500}{\square\micro\metre}
\end{equation} and contains all voxels in thickness direction.
For each of the uncompressed and compressed paper sheets we selected  twelve cutouts to cover a large range of variations in the pore-space morphology. In both cases, the cutouts were chosen such that 
eight of them have a local porosity close to the mean (global) porosity obtained for the entire microstructure data of the sample. Two cutouts represent realizations of particularly dense sheets (smaller porosity) and two further particularly open sheets (higher porosity). To facilitate a later interpretation of the simulation results, all cutouts have been chosen such that their mean thicknesses are approximately equal. Note that one of these cutouts from the compressed sample has not been used in the following analysis, because the CFD-determined porosity of this cutout was implausibly large and possibly indicates an artifact in the mesh generated from the microstructure,  see Section 3.1 below.

\subsection{Transport regime}\label{sec:transport_regime}
We turn to air flow in thickness direction through porous paper sheets under excess pressure. In this transport  regime, we expect that chemical interactions of the molecules in air with the solid matrix are negligible. 
To get laminar flow rather than turbulent flow, the pressure difference between both sides of the sheet must be small enough.
The pressure difference required for the standardized, experimental characterization of volume flows through paper sheets (Gurley test \cite{iso_5636-52013_paper_2013}) is small enough to warrant laminar flow.
Hence, we will use the  pressure difference from the Gurley test in our flow simulations. 
This gives us the opportunity to check whether the simulations are in line with the experimentally obtained air fluxes~\cite{leitl.2023}.
Considering air transport rather than liquid further eases the comparison, because typical liquid-relevant complications such as capillary uptake, wetting in angular pores~\cite{valvatne_predictive_2004, zhao_dynamic_2022}
and associated slip-stick motion~\cite{aslannejad.2017, fischer_four-dimensional_2021} do not appear.

\subsubsection{Navier-Stokes equation.} 
We briefly show how the formulation of the Navier-Stokes equation of laminar transport of gases such as air can be simplified for the use of CFD transport simulations. 
Recall hat the Navier-Stokes equation determines the time-dependent velocity field $u\colon \Omega \times [0,\infty) \to \R^3$ on a given domain $\Omega\subset \R^3$ under the influence of a given pressure field $p\colon \Omega\times  [0,\infty) \to  [0,\infty)$ and negligible gravitational forces for each time $t\geq 0$ and position $x=(x_1,x_2,x_3) \in \Omega$, where 
the domain $\Omega$ contains the (connected) space of pores. 
It is common to add to $\Omega$ additional void volumes above and below the sample, \emph{i.e.}, an inlet and outlet region, such that boundary conditions can be assumed on planar surfaces. 
As the considered pressure difference in our problem is high enough to assume incompressible flow~\cite{bernabe_gaseous_2018}, the Navier-Stokes equation reads
\begin{equation}
\rho \left(\frac{\partial u}{\partial t}  + (u\cdot \nabla) u\right) = -\nabla p + \eta\Delta u\qquad \textup{in } \Omega\times(0,\infty),
\label{eq:NavierStokesIC}
\end{equation}
where  $\eta > 0$ is the dynamic viscosity and $\rho > 0$ is the mass density~\cite{sohr2012navier}. 
At the boundary between the pore and the solid phase, no-slip boundary conditions ensure that the velocity of the gas is equal to that of the rigid solid phase at the wall, which is $0$.
This also ensures that the flow through the pores cannot penetrate the solid matrix. As flows through porous media usually have very small Reynolds numbers, the so-called inertial contribution on the left-hand side of Eq.~\eqref{eq:NavierStokesIC} is neglected. Thus, we assume that 
\begin{equation}
    \left(\frac{\partial}{\partial t}  + (u\cdot \nabla) \right) u = 0\qquad \textup{in } \Omega\times (0,\infty),
\end{equation}
such that the steady-state Stokes equation 
\begin{equation}
 -\nabla p + \eta\Delta u = 0
 \label{eq:Stokes}
\end{equation}
remains.
Then, CFD simulations are performed to determine a solution of Eq.~\eqref{eq:Stokes} together with the continuity equation 
\begin{equation}
 \nabla \cdot u  = 0
 \label{eq:contEq}
\end{equation}
for a given pressure difference at the boundaries to obtain the velocity field $u$ and the local pressure, see Section~\ref{sec:cfd} below.
The volume flow rate $\partial V/\partial t$ through a plane $A\subset\R^3$ that intersects the pore space $\Omega$ is then given by
\begin{equation}
    \frac{\partial V}{\partial t} = \int_{A \cap \Omega} \langle n_A, u\rangle \,\textup{d}\mathcal{H}_2,
    \label{eq:CFD_VFR}
\end{equation}
where $n_A$ is the surface normal vector of $A$ pointing in the direction of the flow, $\langle \cdot ,\cdot \rangle$ denotes the scalar product between two vectors, and $\mathcal{H}_2$ is the two-dimensional Hausdorff measure. Note that we consider the steady state, in which the flow rate $\partial V/\partial t$ does not depend on the time $t$. In the case of the total flow rate through a paper sheet in thickness direction, the volume flow rate is typically determined at a plane that is oriented normally to the pressure difference and is placed in the outlet region. 

\subsubsection{Hagen-Poiseuille equation}
Within the pore network model, the pore space is represented by a graph, in which connections between neighboring pores are geometrically modeled by a cylinder. The volume flow rate of every such connection is determined individually by assuming steady-state, laminar flow through a long, narrow pipe~\cite{xu2022pore, gombosi_gaskinetic_1994,zhao_improved_2020}. In this setting, the flow rate can be modeled using the Hagen-Poiseuille equation~\cite{hutten.2015}. This yields that
\begin{equation}\label{eq:hagen-poiseuille}
    \frac{\partial V}{\partial t}= \frac{\pi R^4}{8\eta L} \Delta p,
\end{equation}
where $\Delta p$ denotes the pressure difference between the two ends of the pipe, $\eta>0$ is the dynamic viscosity, $L>0$ is the length of the pipe and $R>0$ is the pipe radius.
Note that the volume flow rate $\partial V/\partial t$ of a given segment is directly proportional to the difference in pressure between the beginning and the end of the segment. 
Finally, the flow through the entire network must obey a mass balance equation at every vertex. 
For more details on how the graph is constructed in the pore network model and how the linear factors in Eq.~\eqref{eq:hagen-poiseuille} are determined, see Section~\ref{sec:pnm}.

\section{Methods to simulate transport}\label{sec:transp_methods}
We now provide more detailed information on the two approaches used to simulate volume air flow through paper sheets.
In Section~\ref{sec:cfd}, we describe the CFD simulation procedure, and compare the resulting values to experimental data. In Section~\ref{sec:pnm}, the pore network model is described in multiple steps. First, Section~\ref{sec:construct-pnm} describes the construction of the graph representing the pore space morphology. Then, in Section~\ref{sec:flow-pnm}, we show how the graph is used for volume air flow simulations. Section~\ref{sec:conduit-pnm} discusses the impact of the chosen conduit shape used to approximate the volume between two neighboring pores within the pore network model.

\subsection{CFD  simulations}\label{sec:cfd}
The ANSYS code was used to numerically solve the Stokes equation (see Eq.~\eqref{eq:Stokes}) in the pore space of the paper sheets as described in Section~\ref{sec:transport_regime}, where
the simulations were prepared as follows~\cite{leitl.2023}. 
The microstructure of each cutout was provided as a stack of 2D binary images containing voxels corresponding either to the pore or to the solid phase. This volume data was triangulated to obtain the surface mesh of the pore space,
since this surface mesh encloses the pore volume in which the Stokes equation will be solved.
To arrive at a good compromise between high resolution and uniform coverage of surface triangles, we stretched the volume data in thickness direction by a factor of five, loaded this stretched volume data as images in Fiji~\cite{schindelin.2012} to perform the triangulation, and stretched the resulting surface mesh by a factor of 0.2 in thickness direction.
This is done in order to compensate for the difference in lengths between thickness and lateral direction, and arrive at a surface triangulation whose typical side lengths are more proportional to the difference between thickness and lateral size.
As ANSYS relies on a finite volume approach, the pore volume enclosed by the surface mesh was discretized using Numeca Hexpress Hybride.
More details can be found in~\cite{leitl_modelling_2020}.
The surface meshing closes all pores that are not connected to the exterior, \emph{i.e.}, either to an inlet or outlet.
Air permeance was obtained for boundary conditions that correspond to the standardized test to assess the porosity of paper~\cite{iso_5636-52013_paper_2013}. 
The pressure at the top surface was $p_{\mathrm{inlet}}=1.22\,\si{\kilo\pascal}$, and $p_{\mathrm{outlet }}=0$ at the bottom surface.
At the remaining side walls, symmetry boundary conditions are used.
Also the external conditions were chosen to match the standardized conditions. 
The temperature was $T=298\,\si{\kelvin}$, so that the density of air was $\rho_\mathrm{air}=1.184\,\si{\gram\per\cubic\metre}$, the specific heat capacity was $C_p=1006.43\,\si{\joule\per\kilo\gram\per\kelvin}$, and the dynamic viscosity was $\eta=1.838 \cdot 10^{-5}\,\si{\pascal}\si{\second}$. 
For each cutout, the CFD simulations yield the spatially
resolved air pressure and air velocity, and at the outlet side the volume flow rate $\partial V/\partial t$ (via Eq.~\eqref{eq:CFD_VFR}) and the volume flux per unit area $v_{\textup{CFD}}$,  which is given by
\begin{equation}
	\label{eq:meanv}
	v_{\textup{CFD}} = \frac{1}{A_\mathrm{seg}}\frac{\partial V}{\partial t},
\end{equation} 
where $A_\mathrm{seg}$ is defined as in Eq.~\eqref{eq:A_seg}.

\begin{figure}[h!]
	\begin{center}
		\includegraphics[width = .45\textwidth]{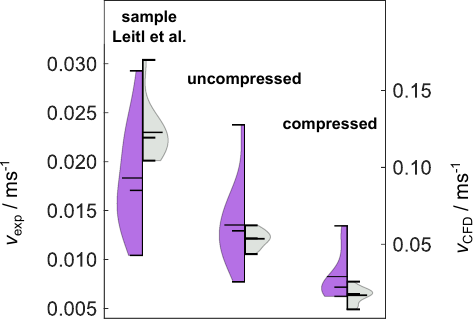}		
	\end{center}
	\caption{Violin plots to compare the fluxes obtained experimentally by the Gurley method (purple, left axis) and the fluxes obtained from CFD simulations (gray, right axis) for a paper sample from~\cite{leitl.2023}, uncompressed, and compressed paper. To ease the comparison, plots of corresponding fluxes are superimposed at the common symmetry axis of the violin plot and a half of each violin plot is hidden. In each plot, the short horizontal bars represent the median and the quartiles, and the long horizontal bar the mean.}\label{fig:CompCFDGurley}
\end{figure}

Figure~\ref{fig:CompCFDGurley} compares the volume fluxes per unit area computed in this way with the fluxes $v_{\mathrm{exp}}$ obtained from experiments for uncompressed (center) and compressed paper (right). In the experiment, a standard volume of \SI{100}{\milli\liter} air is pressed through a paper sheet of a standardized area of \SI{645.2}{\milli\meter\squared} at different positions in the sheet and from top to bottom and from bottom-to-top direction. 
As the measured area of this experiment largely exceeds the area of the cutouts in the CFD simulations, the variations seen in the experiment are much smaller than in the simulations.
The fluxes $v_{\mathrm{exp}}$ and $v_{\mathrm{CFD}}$ differ by a scaling factor between four and five, which is consistent with a previous comparison for a paper made from the same pulp and a porosity comparable to the uncompressed sample~\cite{leitl.2023}, whose simulated and experimentally obtained fluxes are  shown for comparison in the left part of Figure~\ref{fig:CompCFDGurley}.
However, the trends seen in the experiments are captured by the simulations for both samples, in terms of  mean values and variations. 
Hence, in the context of the present study, we consider the CFD-calculated fluxes as the ground truth. They will serve as a reference for the pore network simulations.

\subsection{PNM simulations}\label{sec:pnm}

The PNM approach seeks to predict transport in a geometrically simplified, graph-like representation of the pore space. 
The pore network representation of a pore space contains the positions and sizes of distinguishable pore regions, and the positions and sizes of the cross-sectional contact areas between neighboring regions.

In the network, each vertex represents the center of a pore region. Vertices corresponding to centers of connected pore regions receive an edge in the graph.
A pore space analysis step, described in detail in Section S2 of the Supplementary Information, determines and labels the pore regions and augments the vertices in the graph with information such as the position (center of gravity) and volume of the original pore, and the  diameter of the largest sphere that can be inscribed into the pore. Each edge is marked with information related to the ``bottleneck'' between two connected pores, \emph{i.e.}, the position and extension of the contact area between neighboring pore regions.

The transport between two vertices that are connected by an edge is modeled by solving a transport problem on an approximating conduit shape that is constructed based on the geometric information of each vertex and edge mentioned above.
Correspondingly, the transport simulation is approximated in two steps: (i) determine a pore network representation of the pore space and (ii) perform simulation of Stokes flow on neighboring pores by use of approximating conduit shapes. These steps are explained in Sections~\ref{sec:construct-pnm} and~\ref{sec:flow-pnm}, respectively. Section~\ref{sec:conduit-pnm} discusses the impact of the  conduit shape on the fluxes predicted by PNM.

\subsubsection{Constructing the pore network graph}\label{sec:construct-pnm}

\begin{figure}[h]
	\begin{center}
		\includegraphics[width = 0.65\textwidth]{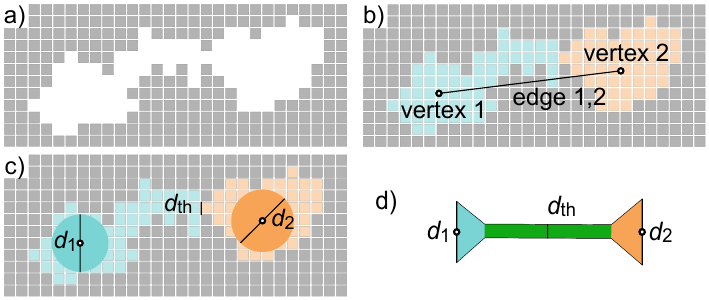}		
	\end{center}
	\caption{Pore space before (a) and after (b) partitioning in distinct pore regions, where each pore region is represented by a vertex, and edges  mark adjacent, connected pore regions.  Analysis of the pore regions provides the positions of the pores, the diameters $d_1$ and $d_2$ of inscribed spheres of maximum diameter, the positions in which the pore regions touch and the largest Euclidean distance $d_{\mathrm{th}}$ therein (c).  Example for a straight pore-throat-pore conduit between two connected vertices in the pore network (d),  where the conduit consists of the pores simplified as truncated cones, whose inlet and outlet diameters are given by $d_1$ and $d_2$, respectively, and a central cylinder with  diameter $d_{\mathrm{th}}$.
    }\label{fig:pore2network}
\end{figure}

The pore network of each cutout was determined with the SNOW algorithm as implemented in the python package \textsf{PoreSpy}~\cite{gostick_porespy_2019}. 
First, the SNOW algorithm divides the pore space of the  segmented 3D image into non-overlapping regions, see Figures~\ref{fig:pore2network}a--\ref{fig:pore2network}c. 
Subsequently, an undirected geometric graph $\mathcal{G}=(\mathcal{V},\mathcal{E})$ is constructed, where each vertex $\mu\in\mathcal{V}\subset\R^3$ corresponds to the centroid of a region in the segmented image, see Figure~\ref{fig:pore2network}b. Then, edges $e=\{\mu,\nu\}\in \mathcal{E}$ are  added between vertices $\mu$ und $\nu$ corresponding to neighboring regions, \emph{i.e.}, between those that share a common boundary.

These vertices and edges receive  geometric information deduced from the tomographic image data of the pore regions. The  volume of a pore region is the total volume of all voxels in a region. The associated pore radius is the maximum value of the Euclidean distance map \cite{soille.2003,maurer.2003}  within each pore region, \emph{i.e.}, it corresponds to the radius of the largest possible sphere that is fully contained in the pore region. The pore surface area is given by the number of solid-pore interface voxels of the region multiplied by the area of a single voxel face. Each edge receives a diameter that corresponds to the largest Euclidean distance found in the cross-sectional area of two adjacent pore regions and a length, that corresponds to the Euclidean distance between the two connected pore centroids~\cite{gostick2017versatile}, see Figure~\ref{fig:pore2network}c.
A more detailed description of how the pore network graph was determined is given in Section~\ref{sec:app-porenetwork} of the Supplementary Information.

\subsubsection{Flow simulation on the network}\label{sec:flow-pnm}
Based on the pore network graphs stated above, the \textsf{openPNM}~\cite{gostick2016openpnm} package was used to determine the stationary air flow for the same excess pressure and external conditions as used for the CFD  simulations~\cite{iso_5636-52013_paper_2013}.
The flux through the network has to obey the mass balance equation at every vertex $\mu\in\mathcal{V}$ in the graph~\cite{ioannidis_network_1993}, \emph{i.e.}, for the net flow rate $(\partial V/\partial t)_{\mu}$ through vertex $\mu$ it holds that
\begin{equation}
	\left(\frac{\partial V}{\partial t}\right)_{\mu} = \sum_{\nu\in N_\mu}\left(\frac{\partial V}{\partial t}\right)_{\mu \nu} = 0,
    \label{eq:PNMlocalMassCons}
\end{equation}
where $N_\mu=\{\nu\in\mathcal{V}\colon \{\mu,\nu\}\in\mathcal{E}\}$ is the set of neighbors of vertex $\mu$ and $(\partial V/\partial t)_{\mu\nu}$ is the flow rate between adjacent vertices $\mu$ and $\nu$, which is positive when the flow is directed towards the vertex $\mu$, and  negative when the flow leaves the vertex $\mu$.

We assume here that the gas behaves like an incompressible liquid with a constant mass density. 
The disregard of compressibility has been shown to have little impact provided that the excess pressure driving the flow is high enough~\cite{bernabe_gaseous_2018}.
Hence, conserving mass is equivalent to conserving volume.

The flow between adjacent vertices $\mu, \nu\in\mathcal{V}$ is driven by the difference in pressures
$p_\mu-p_\nu$ within the pore regions and adopts a form inspired by the Hagen-Poiseuille equation for laminar flow in pipes~\cite{xu2022pore, gombosi_gaskinetic_1994,zhao_improved_2020} given by
\begin{equation}
	\left(\frac{\partial V}{\partial t}\right)_{\mu \nu} = g_{\mu \nu}(p_\mu - p_\nu),
        \label{eq:DarcyLaw}
\end{equation}
where $g_{\mu \nu}>0$ is the local conductivity of the conduit between the adjacent vertices $\mu$ and $\nu$. 

\begin{figure}[h!]
	\begin{center}
		\includegraphics[width = .25\textwidth]{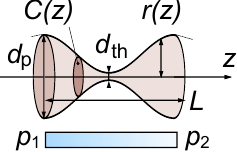}		
	\end{center}
	\caption{Axial-symmetric conduit with varying radius $r(z)$ and cross-section $C(z)$ along the $z$-axis. The inlet is at pressure $p_1$ and the outlet at $p_2$, respectively. The diameter of the inlet and the outlet corresponds to the diameter $d_p$ of the inlet and outlet pore, respectively. The smallest diameter is given by the extension of the bottleneck $d_{\mathrm{th}}$ between inlet and outlet pore.
    }\label{fig:conduit_shapes}
\end{figure}

Note that the local conductivity depends on the size and  the geometric shape of the conduit.
More precisely, the local conductivity (denoted by $g$ in this paragraph) quantifies the ease with which a gas or fluid can pass through a conduit.
For the hydraulic conductance considered here, $g$ is fully determined by the shape of the conduit and the dynamic viscosity $\eta$. Figure~\ref{fig:conduit_shapes} illustrates a general, axial-symmetric conduit with a flow along the $z$-axis. 
We assume that the conduit is placed within the positive quadrant of the coordinate system, with one of its sides aligned with the $x$-$y$-plane, so that its extension along the $z$-axis is from $0$ to $L$ for some length $L>0$.
For each $z\in [0,L]$, let $C(z) \subset \R^3$ be the circular conduit cross-section perpendicular to the flow axis, and $\mathcal{H}_2(C(z))$ the area of $C(z)$, given by means of the two-dimensional Hausdorff measure~$\mathcal{H}_2$. Then, it holds that 
\begin{equation}
	\frac{1}{g} =  16 \pi^2 \eta \int_{0}^{L} \frac{I_p^*(z)}{\mathcal{H}_2(C(z))^2} \,\textup{d}z, 
    \label{eq:hydraulicCond}
\end{equation}	
where $I_p^*(z)$ is the specific polar moment of inertia, defined as  integral over the cross-section $C(z)$ by
\begin{equation}
	I_p^*(z) = \frac{1}{\mathcal{H}_2(C(z))} \int_{C(z)} \Vert x\Vert^2 \,\textup{d}\mathcal{H}_2(x),
\end{equation}
where $\Vert x\Vert$ denotes the Euclidean norm of $x=(x_1,x_2)\in\R^2$.
In Eq.~\eqref{eq:hydraulicCond}, both $\mathcal{H}_2(C(z))$ and $I_p^*$ essentially depend on the geometry of the conduit. Thus, 
analytical descriptions of the shape of the cross sections along the flow axis often allow to derive analytical expressions for the conduit flow~\cite{sochi_newtonian_2013,akbari_viscous_2011}. More details on this are given in Section~\ref{sec:app-conduit} of the Supplementary Information.

In this study, we assume that the  conduit shape is the same throughout the network, where we choose the shape shown in Figure~\ref{fig:pore2network}d, \emph{i.e.}, a cone-cylinder-cone conduit consisting of three segments. 
The inlet segment starts from the pore center with opening diameter $d_1$ and narrows in a conical fashion until the smallest diameter $d_{\textup{th}}$ is reached (pore). A second (straight) segment of cylinder shape with diameter $d_{\textup{th}}$ (throat) follows and connects to a conical-shaped outlet segment with smallest diameter $d_{\textup{th}}$ and final diameter $d_2$.
Note that conduit shapes can be selected from a wide range of shapes as long as  local conductivity through the conduit can be provided to satisfy Eq.~\eqref{eq:DarcyLaw}, see~\cite{zhao_improved_2020,sochi_newtonian_2013,akbari_viscous_2011, miao_new_2017} for examples. A more detailed discussion of the choice of conduit shape will be given in Section~\ref{sec:conduit-pnm} below.

For a conduit between connected pores (represented by the vertices $\mu,\nu\in\mathcal{V}$)  consisting of multiple elements, such as the one shown in Figure~\ref{fig:pore2network}d,  the local conductivity $g_{\mu \nu}$ is obtained
 from the conductivities of the individual conduit  elements~\cite{rodriguez2023predicting}, \emph{i.e.}, 
\begin{equation}
	\frac{1}{g_{\mu \nu}}=\frac{1}{g_{P,\mu}}+\frac{1}{g_{\textup{th},\mu \nu}}+\frac{1}{g_{P,\nu}},
\end{equation}
with $g_{P,\mu}$ and $g_{P,\nu}$ being the conductivities associated to the pore volumes (half-cones in Figure~\ref{fig:pore2network}d), and $g_{\textup{th},\mu \nu}$ is the conductivity of the throat, where all three conductivities  $g_{P,\mu}, g_{P,\nu}$  and $g_{\textup{th},\mu \nu}$ are determined by Eq.~\eqref{eq:hydraulicCond}.

With the local conductivities $g_{\mu \nu}$ in hand, Eq.~\eqref{eq:PNMlocalMassCons} defines a system of linear equations that is solved for the pressure $p_{\mu}$ in each pore with the boundary conditions
$p_{\mathrm{inlet}}=1.22\,\si{\kilo\pascal}$ at all inlet pores and $p_{\mathrm{outlet}} = 0$ at all outlet pores.
Once the pressures $p_{\mu}$ are determined, Eq.~\eqref{eq:PNMlocalMassCons} provides the flow rates through each conduit. 
Then, the total flow rate through the network is the sum over all local volume flow rates entering the network at the inlet pores, \emph{i.e.}
\begin{equation}
	\left(\frac{\partial V}{\partial t}\right)_{\textup{inlet}} = \sum_{\mu \in \mathcal{V_\textup{inlet}}} \left(\frac{\partial V}{\partial t}\right)_\mu = \sum_{\mu \in \mathcal{V_\textup{inlet}}}\sum_{\nu\in N_\mu} g_{\mu \nu}(p_\nu-p_{\mu}),
\end{equation}
where $\mathcal{V}_\textup{inlet} \subset \mathcal{V}$ is the set of all inlet vertices,  which are determined as described in Section~\ref{sec:app-porenetwork} of the Supplementary Information. In analogy to Eq.~\eqref{eq:meanv}, 
the volume flux per unit area $\vpnm$ is  given by
\begin{equation}
	\vpnm = \frac{1}{A_{\mathrm{seg}}}\left(\frac{\partial V}{\partial t}\right)_{\textup{inlet}}.
\end{equation}

\subsubsection{Impact of conduit shape}\label{sec:conduit-pnm}
\begin{figure}[h]
	\begin{center}
		\includegraphics[width = .4\textwidth]{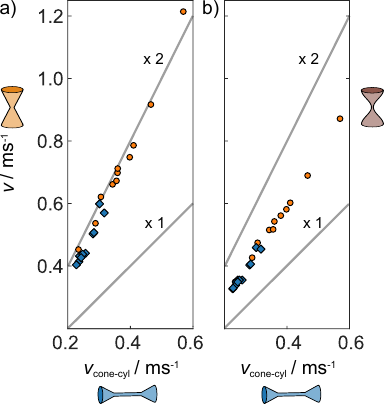}		
	\end{center}
	\caption{Impact of the conduit shapes on the fluxes predicted by PNM for the cutouts of compressed (blue diamonds) and uncompressed (orange circles) paper. The fluxes obtained for half pore-throat-half pore conduits as shown in Figure~\ref{fig:pore2network}d (horizontal axis) are compared to predictions using mirror- and axis-symmetric converging-diverging conduits with diameters $d_{\mathrm{max}} = d_1$ and $d_{\mathrm{min}} = d_{th}$ assuming  a cone (a) and  hyperbolic cosine (b) shape. To ease the comparison, two additional lines corresponding to a direct correspondence $v = v_{\textup{cone-cyl}}$ and a doubling in value $v = 2 v_{\textup{cone-cyl}}$ are inserted.
    } \label{fig:flux_network_poreshape}
\end{figure}

Since the true morphology of the pore space region that connects two different pores is too complex, it must be approximated by using a simplified geometry in PNM to simulate the flow between neighboring pores. As there are many options for doing so, the values of  local conductivities and, hence, of the overall flux $v_{\textup{PNM}}$, depend on the choice of the conduit shape, see Figure~\ref{fig:flux_network_poreshape}. 

To assess the extent to which the predicted flows may vary, we consider the results of our 
simulations using the conduit shape shown in Figure~\ref{fig:pore2network}d as reference and compare the related fluxes to pore network simulations using two other conduit shapes.
The associated conduits assume a converging-diverging shape that are chosen to be mirror-symmetric, \emph{i.e.}, the diameters along the conduit fulfill $d_{\nu,\mathrm{max}} = d_{\mu,\mathrm{max}} = d_{P,\nu}$ and $d_{\mathrm{min}} = d_{\textup{th},\mu \nu}$.  Using the notation $v$ instead of $v_{\textup{PNM}}$,
Figure~\ref{fig:flux_network_poreshape}a shows the flux values of  $v$ obtained for conically shaped conduits for uncompressed (blue diamonds) and compressed  (orange circles) paper,  and compares them to the reference values (denoted by  $v_{\textup{cone-cyl}}$) provided on the horizontal axis.

Although conically shaped conduits tend to show the largest deviations from cylindrical conduits (cf. Figure~\ref{fig:flux_network_poreshape}a), the obtained values of $v$ are practically proportional to the reference values of  $v_{\textup{cone-cyl}}$ with a slope of two.
The same qualitative behavior is obtained for hyperbolic cosine shaped conduits, see Figure~\ref{fig:flux_network_poreshape}b. Also here, the values of  $v$ are proportional to the reference values of $v_{\textup{cone-cyl}}$, but with a reduced slope of 1.5.

The flux through individual conduits of conical and hyperbolic cosine shapes is compared to that of cylindrical conduits in Figure~\ref{fig:flux_poreshape} of the Supporting Material. 

\subsection{Volume flow rates obtained by CFD and PNM simulations}

\begin{figure}[h]
    \centering
        \includegraphics[width = 0.6\textwidth ]{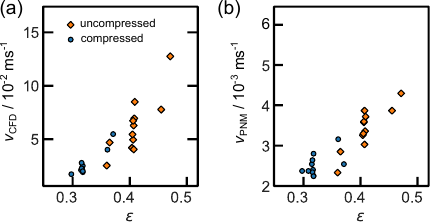}    
    \caption{Volume flux per unit area obtained by CFD  (a) and PNM (b) simulations vs.  local porosity of cutouts for  uncompressed (diamonds) and compressed  (circles) paper.}
    \label{fig:flux_proportionality}
\end{figure}

In this section, we compare the volume flow rates per unit area obtained from CFD and PNM simulations, see Figure~\ref{fig:flux_proportionality}.
The values of the volume flow rates $v_{\mathrm{CFD}}$ and $v_{\mathrm{PNM}}$ differ by more than an order of magnitude. The actual difference in magnitude is determined, at least in part, by the choice of the conduit geometry, as explained above.
Nevertheless, the flows predicted by PNM resemble the trends in the flows obtained by CFD. 
While the fluxes through the uncompressed sample tend to exceed the fluxes in the compressed sample, there is a common pattern regardless of which sample is considered: A higher porosity tends to give a larger flux. For comparable porosities, CFD and PNM simulations predict a marked spread in the flow values, see Figure~\ref{fig:flux_proportionality}.
This spread in flow rates must originate from the details of the pathways realized in each of the cutouts. 
Hence the question arises, whether other microstructure descriptors are capable of explaining this spread because they inherently consider these pathway details. 

\section{Geometric descriptors of pore space morphology}\label{sec:descriptors}
Based on the segmented $\upmu$-CT image data, we can quantitatively analyze the pore space morphology within the measured paper sheets by means of various geometric microstructure descriptors.
In particular, in a previous study~\cite{neumann.2024}, we showed that the differences between uncompressed and compressed paper are heavily reflected in an altered interdependence between pairs of geometric descriptors of the pore space morphology.
As we now aim at linking descriptors that characterize the pore space morphology to the volume flow determined by CFD  simulations, as described in Section~\ref{sec:cfd}, we select an appropriate set of these geometric descriptors. Namely, we consider the porosity $\por$, the specific surface area $S$ of pore space, the mean value $\mu(\tau) $ and standard deviation $\sigma(\tau)$ of geodesic tortuosity of paths through the pore space, and the median radius $\rmax$ of the continuous pore size distribution. It is well known that
each of these descriptors is relevant for transport in porous media~\cite{holzer.2013a,barman.2019, n.2019}. 

\subsection{Computation of geometric descriptors}\label{sec:detdescriptors}
In the following we
briefly explain each of the geometric descriptors considered in the present study and give information on how they can be computed from grid-based 3D image data.
\paragraph{\textbf{Porosity}} The most fundamental and widely used geometric descriptor of  pore space is the porosity $\por\in[0,1]$. Formally, $\por$
is defined as the volume fraction of the pore space, which can be determined by computing  the number of voxels associated with  pores divided by the total number of voxels that do not belong to the background. 
Thus, $\por$ can be computed in a straightforward manner, simply by counting voxels in the image data.
\paragraph{\textbf{Specific surface area}} Another fundamental descriptor is the surface area of pore space. In particular, we consider the specific surface area $S$, \emph{i.e.}, the surface area of pore space per unit volume. 
In order to compute the value of  $S$ from voxelized image data, an algorithm stated in~\cite{schladitz.2006} is used, which is based on local weighted $2\times 2\times 2$ configurations.
\paragraph{\textbf{Geodesic tortuosity}}
The general notion of tortuosity aims to quantify the length of transportation paths through a porous medium in relation to its thickness. Note that there are many different definitions of tortuosity~\cite{Holzer.2023}. However, in the present study, we focus on the concept of geodesic tortuosity. 
This requires the selection of starting and target planes of the image data, for which the geodesic tortuosity shall be computed. 
As we are interested in flow that traverses the paper sample vertically, we chose our starting and target planes as the pore space voxels that belong to the upper- and lower-most layers of voxels in $y$-direction, respectively.
In addition, our goal is to quantify only the lengths of paths that use pores with a certain minimum local volume, so that we ensure that the considered paths contribute to volume flow in a significant way. 
We therefore first determine the pore space that can be filled by spheres of radius $1.5~\unit{\micro\metre}$.
Then, for every pore voxel of the starting plane, a shortest path to the target plane is computed twice by the use of Dijkstra's algorithm~\cite{dijkstra.1959}, where once the path is only allowed to traverse the pore space, while the second time it is also allowed to pass through voxels that belong to the solid phase.
The geodesic tortuosity $\tau\geq 1$ is then defined as the ratio of the lengths of these two shortest paths for every pore  voxel of the starting plane. 
This yields a distribution of values, of which we consider the sample mean $\mu(\tau)$ and sample standard deviation $\sigma(\tau)$.
\paragraph{\textbf{Continuous pore size distribution}} In order to capture the distribution of pore widths across the sample, we consider the continuous pore size distribution $\operatorname{CPSD}\colon [0,\infty)\rightarrow [0,1]$, which is defined using morphological opening~\cite{soille.2003,serra.1982}. More precisely, for each $r\in[0,\infty)$, the value of $\operatorname{CPSD}(r)$ is given by the volume fraction of that part of the pore space that can be covered by spheres of radius $r$ which are completely contained in the pore space. 
To compute $\operatorname{CPSD}(r)$ from voxelized image data,  we consider the subset $P\subset W$ of voxels associated with pores within our sampling window $W\subset\mathbb{Z}^3$. Then, we put
\begin{equation}\label{eq:CPSD_def}
\operatorname{CPSD}(r) =  \frac{\big\vert\big(P\ominus B_{\Z^3}(o,r)\big)\oplus B_{\Z^3}(o,r)\big\vert}{\vert P\vert},
\end{equation}
for any $r\geq 0$, where $\ominus$ and $\oplus$ denote the morphological operations of  erosion and dilation~\cite{soille.2003,serra.1982}, respectively,  $B_{\Z^3}(o,r) = B(o,r) \cap \Z^3$ is the discretized  open ball of radius $r$ centered at the origin, and $\vert B\vert$ denotes the number of elements of any set $B \subset \mathbb{Z}^3.$
Note that the numerator on the right-hand side of Eq.~\eqref{eq:CPSD_def} 
can be efficiently computed by means of the Euclidean distance transform~\cite{soille.2003,maurer.2003}. In particular, we are interested in the median  $r_{\max}$ of  CPSD, which is given by $r_{\max}=\max\{r\ge 0: \operatorname{CPSD}(r)\geq 1/2\}$. 
This quantity is used in the definition of constrictivity, which is a useful descriptor in quantifying the strength of bottleneck effects~\cite{holzer.2013b}. 

\subsection{Correlations between pairs of geometric descriptors}
\begin{figure}[h]
    \centering
    \includegraphics[width=.8\textwidth]{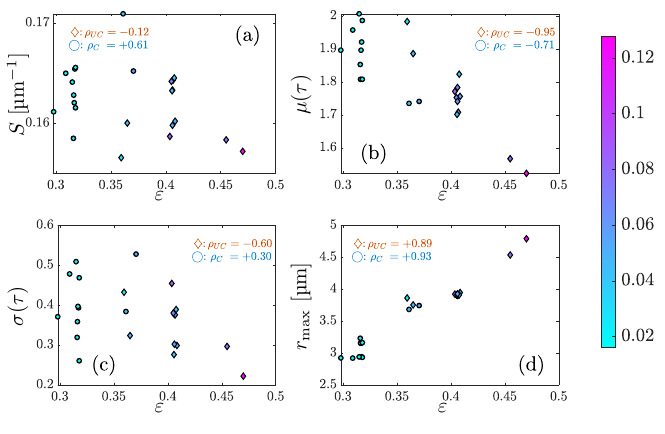}
    \caption{Scatter plots visualizing the interdependence of geometric descriptors for the cutouts of uncompressed (diamonds) and compressed (circles) paper sheets. The color coding indicates the values of the volume flow obtained by CFD simulations. Additionally, the corresponding values of the  Pearson correlation coefficient are displayed on each figure.}
    \label{fig:descriptor_scatter}
\end{figure}

Figure~\ref{fig:descriptor_scatter} shows scatter plots of the results which we obtained for the geometric descriptors stated in Section~\ref{sec:detdescriptors},
for the cutouts of both uncompressed and compressed paper sheets. It also shows the Pearson correlation coefficients for each pair of displayed descriptors, where we can see that some descriptor pairs show a rather  strong correlation. Namely, the pairs $\por$ and $\mu(\tau)$ (Figure~\ref{fig:descriptor_scatter}b) as well as $\por$ and $\rmax$ (Figure~\ref{fig:descriptor_scatter}d) are strongly correlated in both samples. 

In line with this, the associated fluxes, imposed by color in Figure~\ref{fig:descriptor_scatter}, tend to be higher the higher the porosity $\por$ and the median radius $\rmax$, and the lower the mean geodesic tortuosity $\mu(\tau)$, which can be seen best  in Figures~\ref{fig:descriptor_scatter}b and~\ref{fig:descriptor_scatter}d.
In contrast to this, $\por$ and $S$ only show a significant correlation for the cutouts of the compressed sample, see Figure~\ref{fig:descriptor_scatter}a, as also found in our previous work~\cite{neumann.2024}.

Figure~\ref{fig:descriptor_scatter} also shows that at least half of the data points per sample share a practically identical local porosity $\por$. 
Although our pick of cutouts with similar local porosity may adversely affect the estimation of the Pearson correlation coefficient for the interrelations representing the whole sample, it will boost the importance of the other geometric descriptors in explaining variations in the flow. 
Note that the fact that the geometric descriptors considered in the present study are correlated affects the variance and interpretability of the estimated coefficients in the regression models that we will present in Section~\ref{sec:formulas} below. However, the presence of correlation among explanatory variables does not in general influence the ability to obtain a good fit~\cite{kutner.2004}, see also Section~\ref{sec:discussion_separate}.

\section{Structure-property relationships}

In this section, we present six empirically derived regression models that will be used to predict the volume flow $v_{\mathrm{CFD}}$ as determined by CFD simulations, from geometric descriptors of pore space morphology introduced in Section~\ref{sec:descriptors}, or from the volume flow prediction $\vpnm$ of PNM simulations. In the following, the target variable of each regression will be denoted by $v^{(i)}$ for $i\in\{1,\ldots,6\}$, which will be used to predict the value of $v_{\mathrm{CFD}}$.

\subsection{Regression models for structure-property relationships}\label{sec:formulas}

The regression models that we consider are all of power-law type.
Such models are easy to implement, and their predictions do not fundamentally differ from predictions using more complex relations~\cite{Hommel.2018}.
The accompanying disregard of a critical porosity, often needed to describe media of small porosity~\cite{Hommel.2018, gebart_permeability_1992, Nabovati.2009} is justified, as measured and simulated fluxes  never vanished and the considered interval of porosity values is rather narrow, see Figures~\ref{fig:descriptor_scatter}a and~\ref{fig:descriptor_scatter}b.

\subsubsection{Regression models using only geometric descriptors}\label{sec.fiv.one.one}

The simplest and most well-known relationship is given by
\begin{equation}\label{eq:v1}
    \rate{1}(\por) = c_0\por^{c_1},
\end{equation}
for some $c_0,c_1\in\R$,
where only the porosity $\por$ is considered as an explanatory variable. This relation is widely used, as porosity is by far the most accessible geometric descriptor.
With the next regression model, we add further geometric descriptors to capture the pore space morphology in more detail. It is given by 
\begin{equation}\label{eq:v2}
    \rate{2}\big(\por, \mu(\tau), \sigma(\tau), S\big) = c_0\por^{c_1}\mu(\tau)^{c_2}\sigma(\tau)^{c_3}S^{c_4},
\end{equation}
for some $c_0,c_1,c_2,c_3,c_4\in\R$.
We can expect that $\rate{1}(\por)\not=\rate{2}\big(\por, \mu(\tau), \sigma(\tau), S\big)
$, since the distribution of the lengths of transportation pathways through the material has a significant impact on the resulting volume flow rate. We also remark that the specific surface area $S$ is not a dimensionless descriptor such as porosity or geodesic tortuosity. 
In order to analyze the improvement provided by the additional geometric descriptors considered in Eq.~\eqref{eq:v2}, it will be useful to additionally investigate the two simplified models
\begin{equation}\label{vau.two.one}
    \rate{2,1}\big(\por, \mu(\tau), \sigma(\tau)\big) = c_0\por^{c_1}\mu(\tau)^{c_2}\sigma(\tau)^{c_3},
\end{equation}
for some $c_0, c_1, c_2, c_3\in\R$ and
\begin{equation}\label{vau.two.two}
    \rate{2,2}(\por, S) = c_0\por^{c_1}S^{c_2},
\end{equation}
for some $c_0,c_1,c_2\in\R$.

\subsubsection{Regression models including the results of PNM simulations}\label{sec.fiv.one.two}
Power laws in the form presented above are a helpful tool to test which descriptors indicate local intricacies of the pore space geometry that may have been oversimplified in the PNM simulations described in Section~\ref{sec:pnm}.
On the other hand, we can assess the predictive power of the PNM itself with the relation
\begin{equation}\label{eq:v3}
    v^{(3)}(\vpnm) = c_0\vpnm^{c_1},
\end{equation}
for some $c_0,c_1\in\R$. 
This formulation allows for a comparison to $v_{\mathrm{CFD}}$ beyond strictly linear relations and automatically accounts for the difference in scale of the CFD and PNM predictions.
By extending the relation given in Eq.~\eqref{eq:v3}  with geometric descriptors as additional factors, the improvement of the prediction depending on the considered descriptors indicates to which extend the descriptors "correct" the original PNM prediction.
A combination of the models stated in Eqs.~\eqref{eq:v1} and~\eqref{eq:v3} is given by
\begin{equation}\label{eq:v4}
    v^{(4)}(\por, \vpnm) = c_0\por^{c_1}\vpnm^{c_2},
\end{equation}
for some $c_0,c_1,c_2\in\R$. 
Moreover, using the geometric descriptor $\rmax$, we can further extend the relationship given in Eq.~\eqref{eq:v4} and obtain
\begin{equation}\label{eq:v5}
    v^{(5)}(\por, \rmax,\vpnm) = c_0\por^{c_1}\rmax^{c_2}\vpnm^{c_3},
\end{equation}
for some $c_0,c_1,c_2,c_3\in\R$. 
Note that the median radius $\rmax$ and the specific surface area $S$ have a physical unit, which introduces information on the typical length scale of the pore space.
Finally, we consider a combination of Eqs.~\eqref{eq:v2} and~\eqref{eq:v3}, which leads to
\begin{equation}\label{eq:v6}
    v^{(6)}\big(\por, \mu(\tau), \sigma(\tau), S, \vpnm\big) = c_0\por^{c_1}\mu(\tau)^{c_2}\sigma(\tau)^{c_3}S^{c_4}\vpnm^{c_5},
\end{equation}
for some $c_0,c_1,c_2,c_3,c_4,c_5\in\R$. 
The latter relationship contains five explanatory variables and, therefore,  will likely provide the best fit simply by having the largest degree of freedom. Its purpose is primarily to serve as a reference when assessing the improvement of adding a particular descriptor.

\subsection{Fitting and validation of regression models}
We briefly explain the fitting procedure for the models proposed in Section~\ref{sec:formulas} and the criteria that we use to assess the goodness of fit. 
All models that we consider in the present study are power-type models of the general form 
\begin{equation}\label{eq.twe.fiv}
    v = c_0 x_1^{c_1}\ldots x_n^{c_n},
\end{equation}
for some $n\in\{1,\ldots,5\}$, where $x_1,\ldots,x_n\in\R$ are explanatory variables, $c_0,\ldots,c_n\in\R$ are coefficients that need to be determined, and $v\in\R$ is the prediction of  the volume flow determined by CFD  simulations as described in Section~\ref{sec:cfd}.
The explanatory variables $x_1,\ldots,x_n$ are either geometric descriptors as presented in Section~\ref{sec:descriptors} or the volume flow determined by PNM simulations as described in Section~\ref{sec:pnm}.  
Applying the natural logarithm to both sides of Eq.~\eqref{eq.twe.fiv} yields
\begin{equation}\label{eq:linear_model}
    \log(v) = \log(c_0) + \sum_{i=1}^n c_i \log(x_i),
\end{equation}
which is a linear regression problem with explanatory variables $\log(x_1),\ldots, \log(x_n)$, response variable $\log(v)$, and coefficients $\log(c_0),c_1,\ldots,c_n\in \R$. 
We determine the coefficients of this linear regression problem by means of the function $\texttt{fitlm}$ in Matlab~\cite{MATLAB}, which uses least square estimation. 
We then receive a prediction  $v$ of $v_{\mathrm{CFD}}$  by applying the exponential function to the prediction $ \log(v)$ obtained byEq.~\eqref{eq:linear_model}.

In order to quantify the goodness of fit, 
we consider both the coefficient of determination (denoted by $R^2$) and the mean absolute percentage error (denoted by $\textup{MAPE}$). If $m\in\N=\{1,2,\ldots\}$ is the number of data points available for the considered regression, these quantities are defined as
\begin{equation}\label{eq:rsq_and_mape}
    R^2 = 1 - \frac{\sum_{k=1}^{m} \big(\log(v_{\mathrm{CFD},k})-\log(v_k)\big)^2}{\sum_{k=1}^{m}\big(\log(v_{\mathrm{CFD},k}) - \overline{v}_{\mathrm{CFD}}\big)^2} \;\; \textup{and}\;\; \textup{MAPE} = \frac{100}{m}\sum_{k=1}^{m} \left\vert\frac{\log(v_{\mathrm{CFD},k}-\log(v_k)}{\log(v_{\mathrm{CFD},k})}\right\vert,
\end{equation}
where $\log(v_{\mathrm{CFD},1})\ldots,\log(v_{\mathrm{CFD},m})$ are the (logarithmic) ground truth values resulting from CFD simulations as described in Section~\ref{sec:cfd}, $\log(v_1),\ldots,\log(v_m)$ are the corresponding predicted (logarithmic) values of the linear regression given in Eq.~\eqref{eq:linear_model}, and $\overline{v}_{\mathrm{CFD}}$ is the sample mean of $\log(v_{\mathrm{CFD},1})\ldots,\log(v_{\mathrm{CFD},m})$. 

Note that the coefficient of determination $R^2$ aims to quantify how much of the variance in the data is explained by the model fit, where, to ensure the proper interpretability of this quantity, it is crucial that the regression problem considered is linear~\cite{Spiess.2019}.
Although we basically consider regressions of power type, the logarithmic transformation applied in Eq.~\eqref{eq:linear_model} yields a linear regression problem. 
Therefore, we always use logarithmic values to evaluate both $R^2$ and $\textup{MAPE}$. 
That is, the ground truth values $\log(v_{\mathrm{CFD},1})\ldots,\log(v_{\mathrm{CFD},m})$ are given by the logarithms of the simulated volume flows, computed as described in Section~\ref{sec:cfd}, and the predictions $\log(v_1),\ldots,\log(v_m)$ are the predicted volume flows obtained from the linear regression problem described by Eq.~\eqref{eq:linear_model}. 
In this way, we ensure that the resulting values allow for an adequate comparison between the regression models stated in Section~\ref{sec:formulas}. 
However, due to the limited amount of available data, we always use the same data to fit the coefficients $\log(c_0),c_1,\ldots,c_n\in \R$ in Eq.~\eqref{eq:linear_model}  as we do to evaluate the statistics $R^2$ and $\textup{MAPE}$  defined in Eq.~\eqref{eq:rsq_and_mape}.

\section{Results and discussion}
We now present the results we obtained for the fitted regression models stated in Section~\ref{sec:formulas}. At first, all regression models are fitted separately to both datasets of uncompressed and compressed paper sheets, yielding two separate sets of coefficients for each regression model. In this way, we can analyze the predictive power of the involved descriptors in a sample-specific manner, yielding insights into how the dependency structure between geometric descriptors and the volume flow rates obtained by CFD and PNM simulations  changes between different samples.

In Section~\ref{sec:discussion_separate}, we consider regression models that involve only purely geometric descriptors of pore space. Afterwards, in Section~\ref{sec:discuss_PNM}, 
we show how incorporating the results of PNM simulations can be used to improve the predictions of the volume flow rate. 
Finally, in Section~\ref{sec:discuss_jointfit}, we analyze how well the considered regression models generalize across uncompressed and compressed paper sheets, by using only one set of coefficients for each regression model that has been jointly determined from both data sets. The analysis of these regression models, and the comparison to the separately fitted regression models discussed before, will provide insight into how to formulate general relationships that apply to varying grades of paper sheets with different morphologies. The values determined for each of the coefficients $c_0,\ldots,c_n$ of the regression models considered are listed in Tables~\ref{tab:uc_coefficients} -- \ref{tab:joint_fit_coefficients}
of the Supplementary Material.

\subsection{Regression models using only geometric descriptors}
\label{sec:discussion_separate}
\begin{figure}[h]
	\centering
	\includegraphics[width=.8\textwidth]{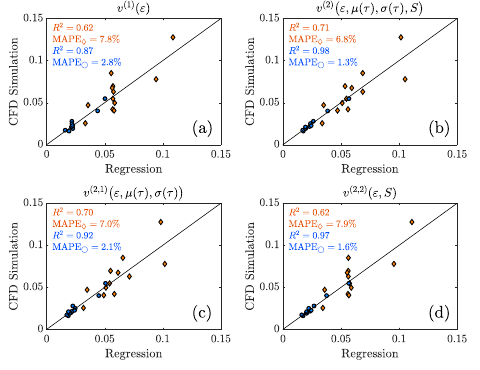}
	\caption{Scatter plots of volume flow rates obtained by CFD simulations vs. the values predicted by the respective regression models. Separate fits of regression models have been determined for the data points of the uncompressed~(orange diamonds) and compressed sample~(blue circles). The coefficient of determination $R^2$ and the mean absolute percentage $\operatorname{MAPE}$ are displayed for both samples on each plot. As a guide to the eye, the black line highlights the diagonal on which the results of CFD simulation and regression coincide.}
	\label{fig:FitDescriptors2}
\end{figure}

In this section, we only consider the purely geometric regression models stated in Section~\ref{sec.fiv.one.one}. These will give insights into predicting volume flow rates by CFD simulations based only on geometric information of the pore space morphology. The results of the fits are visualized in Figure~\ref{fig:FitDescriptors2}, where the predictions of the corresponding regression models are plotted against the flow rates of CFD simulations.

As mentioned above, for each of the regression models, two sets of coefficients $c_0,\ldots,c_n$ have been determined separately for uncompressed and compressed paper sheets. Therefore, Figure~\ref{fig:FitDescriptors2} also shows two values for the coefficient of determination $R^2$ and the mean absolute percentage error $\operatorname{MAPE}$, displayed in orange and blue for the uncompressed and compressed sample, respectively.

We start by considering the simplest model $\rate{1}$ given in Eq.~\eqref{eq:v1}, which attempts to establish a direct connection between porosity $\por$ and  volume flow rate $v_{\textup{CFD}}$ obtained by CFD simulations, see Figure~\ref{fig:FitDescriptors2}a.
While $\rate{1}$ predicts the overall trend quite well, it cannot distinguish samples with a similar porosity, which results in clusters of vertically arranged data points in Figure~\ref{fig:FitDescriptors2}a. 
This effect results in the low value of $R^2=0.62$ in the uncompressed case, but is less detrimental in the compressed sample, where we receive a value of $R^2=0.87$.

The vertically clustered points in Figure~\ref{fig:FitDescriptors2}a clearly indicate that there are factors influencing the volume flow that cannot be explained by porosity alone.
This is consistent with the situation found in similar paper sheets~\cite{leitl.2023}. 
Therefore, our next step is to consider three further descriptors of pore space morphology, that are all incorporated in the model $\rate{2}$
given in Eq~\eqref{eq:v2}.
The first two descriptors relate to the distribution of geodesic tortuosity $\tau$, which enters $\rate{2}$ in the form of its mean value $\mu(\tau)$ and  standard deviation $\sigma(\tau)$, while the third is the specific surface area $S$.
The fit with respect to $\rate{2}$ substantially improves the fit by $\rate{1}$ regardless of the sample, see  Figure~\ref{fig:FitDescriptors2}b. In particular, 
for the compressed sample a value of $R^2=0.98$ and for the uncompressed sample a value of $R^2=0.71$ are achieved.

Obviously, it is not surprising that we achieve an improved fit in both cases when refining the model through additional descriptors. 
However, individual descriptors improve the fit by varying degrees in a sample-specific way. To make this clear, we
consider the regression models $\rate{2,1}$ and $\rate{2,2}$  given in Eqs.~\eqref{vau.two.one} and \eqref{vau.two.two}, which both use a subset of the descriptors considered in $\rate{2}$, see Figures~\ref{fig:FitDescriptors2}c and~\ref{fig:FitDescriptors2}d.
For the uncompressed sample (orange diamonds in Figure~\ref{fig:FitDescriptors2}), the accuracy of  $\rate{2,1}$ is almost identical to that of $\rate{2}$, with a coefficient of determination of $R^2=0.70$ for $\rate{2,1}$,  compared to $R^2=0.71$ for $\rate{2}$. Thus, for the data set of this sample, one can argue that information on the specific surface area $S$  is redundant to predict  the volume flow rate
$v_{\textup{CFD}}$ obtained  
by CFD simulations, while information on the geodesic tortuosity is more relevant. 
On the other hand, for the compressed sample, the regression model $\rate{2,2}$ yields a value of $R^2=0.97$ in comparison to a value of $R^2=0.98$ for the more complex model $\rate{2}$. Thus, in this case, information on the specific surface area $S$ provides an  enormous improvement compared to the regression model $\rate{1}$, while the information on geodesic tortuosity seems redundant.

Intuitively, one might assume that any new descriptor, which is highly correlated with already present descriptors, would not provide enough new information and therefore cannot significantly improve the fit.
However, as shown in~Figure~\ref{fig:descriptor_scatter}a, the porosity $\por$ and the specific surface area $S$ are significantly correlated in the compressed case, with only a low correlation coefficient in the uncompressed case. Nevertheless, the compressed case benefits more from knowledge on the specific surface area $S$, while the uncompressed case sees more improvement by incorporating  information on the geodesic tortuosity.

This illustrates that a descriptor cannot be deemed as redundant or not, just based on its correlations with other  (already present) descriptors.
Note that the presence of correlations among explanatory variables affects the variance and interpretability of the values of the estimated coefficients $c_0,\ldots,c_n$
provided in Tables~\ref{tab:uc_coefficients} -- \ref{tab:joint_fit_coefficients}. 
In particular, we cannot formulate a quantitative statement that a certain increase of an explanatory variable would have a certain effect on the predictor variable, as it might not be possible to increase an explanatory variable without affecting the others. 
However, these correlations do in general not influence the quality of the obtained fit~\cite{kutner.2004}. In turn, the absence of correlations among explanatory variables does not generally guarantee an improved fit.
In many machine learning approaches, feature selection is performed according to a so-called maximum-relevance-minimum-redundancy criterion, which aims to weigh the correlation between a given feature and the target observation against the correlation between the given feature and other, already present, features~\cite{peng_feature_2005, van_der_linden_machine_2016}. However, such a systematic approach requires an accurate estimation of the probability distributions of involved features, which is not feasible in our case due to the small data base. Instead, we follow the same idea by manually evaluating our metrics for the goodness of fit and analyzing the correlation between the present descriptors.

\subsection{Combining geometric information with PNM simulations}\label{sec:discuss_PNM}
We now consider the regression models $\rate{3}$  to $\rate{6}$ stated in Section~\ref{sec.fiv.one.two}, which all involve the volume flow rate $\vpnm$ as determined by PNM simulations. The model $\rate{3}$ uses only $\vpnm$, while $\rate{4}$, $\rate{5}$ and $\rate{6}$ combine information from geometric descriptors and PNM simulations. The analysis of these models provides insight about the accuracy of predicting the results of  CFD simulations based on those of PNM simulations, and whether it is possible to find geometric descriptors of pore space morphology that mitigate the discrepancies between the two predictions. Similarly to the previous section, all the models considered use two sets of coefficients $c_0,\ldots,c_n$, which have been estimated separately based on the data of uncompressed and  compressed paper sheets, respectively.

\begin{figure}[h]
	\centering
    \includegraphics[width=.8\textwidth]{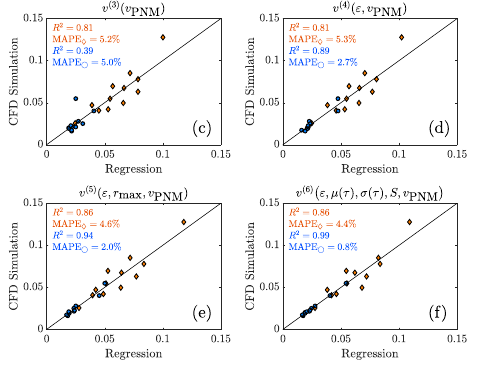}
	 \caption{Scatter plots of volume flow rates obtained by CFD simulations vs. the values predicted by regression models $\rate{3}$ to $\rate{6}$ (panels ~\ref{fig:regression_corrPNMs_fit3}a  to~\ref{fig:regression_corrPNMs_fit3}d) aiming at correcting the flow rate predictions obtained by PNM simulations. Separate fits of the regression model to the uncompressed~(orange diamonds) and compressed sample~(blue circles). The coefficient of determination $R^2$ and the mean absolute percentage $\operatorname{MAPE}$ are displayed for both samples on each plot. As a guide to the eye, the black line highlights the diagonal on which the results of CFD simulation and regression  coincide.
    }
	\label{fig:regression_corrPNMs_fit3}
\end{figure}

Figure~\ref{fig:regression_corrPNMs_fit3}a shows how well flow rates obtained from PNM simulations predict the CFD flow values, using the regression model $\rate{3}$. 
The overall fit is clearly better for the uncompressed case than for the compressed case, with a higher $R^2$-value and lower $\textup{MAPE}$. 
However, the lack of accuracy in the compressed case ($R^2 = 0.39$) is mainly due to an outlier associated with a CFD volume flow of 
$v_{\textup{CFD}} =$ \SI{0.05}{\metre / \second}, see  Figure~\ref{fig:regression_corrPNMs_fit3}a.

The regression model $\rate{4}$ considers porosity as an additional descriptor that can correct flow rates determined by PNM simulations.  
With the additional descriptor, the model can correct for the above mentioned outlier, increasing the coefficient of determination from $R^2=0.39$ to $R^2=0.89$ in the compressed case, see Figure~\ref{fig:regression_corrPNMs_fit3}b.
Furthermore, a comparison between the fits of $\rate{4}$ and $\rate{1}$ also shows that incorporating PNM simulations already significantly improves the classical approach of predicting volume flow rates only by means of porosity.

The regression models $\rate{5}$ and $\rate{6}$ reveal, in analogy to the transition from $\rate{1}$ to $\rate{2}$, which geometric descriptors in addition to $\por$ significantly correct the PNM-predicted flow rates and whether a strong correlation of a descriptor with porosity (or the absence thereof) controls the extent of improvement.
Recall that $\rate{5}$ considers the porosity $\por$ and the median pore radius $r_{\mathrm{max}}$, obtained from the continuous pore size distribution. Here we observe a similar phenomenon as in the previous section.
The median pore radius $r_{\mathrm{max}}$ is strongly correlated with porosity $\por$ for both samples, see Figure~\ref{fig:descriptor_scatter}d.
Despite this strong correlation, the model provided by $\rate{5}$ fits the data of the CFD volume flows significantly better than $\rate{4}$ for both samples, see Figure~\ref{fig:regression_corrPNMs_fit3}c. 
Finally,  $\rate{6}$ considers all geometric descriptors, that were already used in $\rate{2}$ as explanatory variables,  in addition to the PNM-predicted flow rate. 
In comparison to $\rate{2}$, the additional information of $\vpnm$ still leads to an improvement of an already highly accurate prediction, \emph{i.e.,} from an $R^2$-value of $0.98$ to $0.99$, see Figure~\ref{fig:regression_corrPNMs_fit3}d. 
Most notably, the models $\rate{5}$ and $\rate{6}$ yield nearly identical precisions in the uncompressed case, although $\rate{6}$ uses the three descriptors $\mu(\tau)$, $\sigma(\tau)$, and $S$ to replace only the one descriptor $\rmax$ in $\rate{5}$. Yet again, this is in contradiction to our expectations, given that the porosity $\por$ and the median pore radius $\rmax$  have a high  correlation coefficient of $\rho=0.89$ in the uncompressed case.

\subsection{Generalizing relationships across different paper grades}\label{sec:discuss_jointfit}

With the regression models at hand, the question arises whether the models would also satisfactorily predict the CFD volume flow rates for a wider range of microstructures.  
To test this, we now determine a single set of coefficients for each of the regression models stated in Section~\ref{sec:formulas}, which is jointly determined from all data points of both the combined uncompressed and compressed samples.
As we want to compare the performance of these jointly fitted models to the separately fitted models of the previous sections, we also require the metric for the goodness of fit to be comparable. Therefore, we again determine the values of the coefficient of determination $R^2$ and the mean absolute percentage error $\operatorname{MAPE}$ for each regression. More precisely, we evaluate these metrics once only on the data points of the uncompressed sample and once only on the compressed sample. In this way, we obtain a value of $R^2$ and $\operatorname{MAPE}$ for each sample and each of the jointly fitted models that are directly comparable to the $R^2$- and $\operatorname{MAPE}$-values of the separately fitted models. 

\begin{figure}[h]
\centering
\includegraphics[width = 0.65\textwidth ]{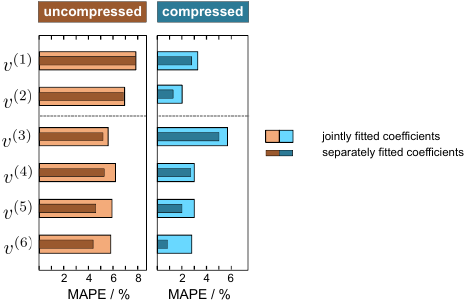}   

	\caption{Mean average percentage error (MAPE) between CFD-predicted volume flow rates and predictions of the regression models $\rate{1},\ldots,\rate{6}$, each fitted to the combined data points of uncompressed and compressed paper (wide bars), or separately to 
    uncompressed paper and compressed paper, respectively (narrow bars).}
	\label{fig:joint_MAPE}
\end{figure}

Figure~\ref{fig:joint_MAPE} collects the $\operatorname{MAPE}$-values for all jointly fitted and separately fitted models, evaluated for both data sets of  the uncompressed and the compressed sample. 
Bars with a light shading indicate the $\operatorname{MAPE}$-values of the jointly fitted model, while the narrow bars with darker shading show the $\operatorname{MAPE}$-values from the previous sections, in which separate sets of coefficients $c_0,\ldots,c_n$  were used for each sample. 
Corresponding scatter plots for these regression models are shown in Figure~\ref{fig:regression_joint_fit_allmodels} of the Supplementary Information.

The models $\rate{1},\ldots,\rate{6}$ considered in Figure~\ref{fig:joint_MAPE} are ordered with increasing complexity (from top to bottom): Regression models $\rate{1},\rate{2}$ exclusively contain geometric descriptors of the pore space, while regression models $\rate{3},\ldots,\rate{6}$ are associated with comparison and corrections of the predictions by means of PNM simulations. 
Obviously, with increasing complexity of the models, the values of $\operatorname{MAPE}$ are decreasing. However, we can see that in some cases the added complexity leads only to an improved model if it can specialize to certain types of microstructures.  For example, this is the case for $\rate{6}$ in the compressed case,  see Figure~\ref{fig:joint_MAPE}.
In general, the more complex the model, the greater the discrepancy between the separately fitted models and the jointly fitted model.
In particular, we see that for both samples, the jointly fitted models $\rate{4},\ldots,\rate{6}$ all show very similar performance, indicating that the additional complexity of $\rate{5}$ and $\rate{6}$ can only be capitalized on when fitting to specific data sets, but not when generalizing across different grades of paper. However, the decrease in MAPE between $\rate{1}$ and $\rate{4}$ is consistent between both samples, indicating that combining the results of the PNM simulations with geometric descriptors yields an improvement regardless of the specific paper grade considered.

In general, using porosity $\por$ as a correction factor in model $\rate{4}$ for the prediction obtained by PNM simulations helps to better explain the flow rates in the compressed sample, \emph{i.e.}, the denser sample. The descriptor $\rmax$ also has a significant impact for the compressed sample.  
However, it remains unclear whether this is due to sample-specific details of the pore network model. To decide this, the structures harboring local flows need to be resolved in more detail. 
In such a local analysis, the structure of the pore network model could also play a role, as PNM simulations on conduits of the same shape tend to neglect too many details of the actual shape of the interface formed between connected pores~\cite{zhao_improved_2020}. 

\section{Summary and Conclusion}

In this study, the volume fluxes of air through various cutouts of two different samples of porous paper sheets were determined by using CFD and PNM simulations. The fluxes from CFD simulations serve as ground truth in our study, while PNM simulations provide a simplified and computationally much less demanding alternative. 
In addition to volume fluxes, various geometric descriptors of the pore space have been determined for each of the cutouts based on the tomographic $\upmu$-CT image data.
In particular, we focused on analyzing the performance of the PNM simulations in comparison to CFD simulations and investigating whether additional information on the pore-space morphology through geometric descriptors can be used to improve the fluxes predicted by PNM simulations.
We also directly analyzed structure-property relationships between geometric descriptors and volume fluxes of CFD simulations.
This was done by fitting six different regression models of power-law type to the data obtained for the cutouts of both samples. These regression models were fitted separately to the two data sets of compressed and uncompressed paper samples to analyze the performance of the models in a sample-dependent way. 
Subsequently, we also fitted the models to the joint data set of both samples to see how well the individual models generalize across both samples.

Our analysis showed that, while volume fluxes determined by PNM simulations qualitatively resemble the fluxes determined by CFD simulations, involving additional geometric descriptors significantly improves the accuracy of the predictions. 
From the sample-specific analysis, we see that the ideal choice of geometric descriptors depends on the chosen sample, which is in line with previous results showing altered correlation structures of geometric descriptors in paper sheets after compression~\cite{neumann.2024}.
However, even simply including porosity in the regression model already yields an improvement for both samples.
In particular, the regression model using only porosity and PNM fluxes in some cases outperforms more complex relationships that directly model the CFD fluxes by multiple geometric descriptors of pore space, which underlines the potential of PNM as a more cost-efficient simulation tool. 

\bibliography{Literatur}
\bibliographystyle{unsrt}
\section*{Statements and Declarations}
\subsection*{Funding}
	
The authors appreciate financial support from the Christian Doppler Research Association, the Federal Ministry for Digital and Economic Affairs, and the National Foundation for Research, Technology, and Development, Austria. M.N. acknowledges funding by the German Research Foundation (DFG) under Project ID 390874152 (POLiS Cluster of Excellence, EXC 2154). K.Z. thanks the "TU Graz Lead Project LP-03: Porous Materials @ Work for Sustainability" for inspiration.
	
\subsection*{Competing interests}
The authors have no relevant financial or non-financial interests to disclose.
\subsection*{Author contributions}
	E.M.C., R.S., A.H. and I.M. acquired the micro-CT scans. M.N., E.M.C. and A.H. segmented the image data. U.H. suggested use cases and supported the interpretation of the model predictions.
	M.N. and P.G. performed the statistical analysis of the microstructure data. 
    A.R. and P.L. performed the CFD simulations.
    E.B. and M.F. performed the PNM simulations. M.N., P.G. and K.Z. conceptualized and wrote the manuscript.  M.N., P.G., V.S. and K.Z. were involved in the editing of the manuscript.  
	All authors have read and approved the final version of the manuscript.
	\subsection*{Data Availability}
    Additional data is available as supplementary information.
    The datasets generated during and/or analyzed during the current study are available from the corresponding author on reasonable request.
\clearpage
\setcounter{figure}{0}
\setcounter{equation}{0}
\setcounter{table}{0}
\renewcommand{\theequation}{S\arabic{equation}}
\renewcommand{\thetable}{S\arabic{table}}  
\renewcommand{\thefigure}{S\arabic{figure}}

\renewcommand*{\thesection}{S}
\section*{Supplementary Information}
\subsection{Fitted regression coefficients}
We provide the values of the coefficients of each descriptor determined for the regression models $v^{(1)},\ldots,v^{(6)}$ discussed in the main text. 
Tables~\ref{tab:uc_coefficients} and~\ref{tab:c_coefficients} show the coefficients of the fits based on the two separate data sets of uncompressed and compressed paper sheets, respectively, as discussed in Section~\ref{sec:discussion_separate}. Table~\ref{tab:joint_fit_coefficients} shows the coefficients of the fits based on the joint data base of both uncompressed and compressed paper sheets as discussed in Section~\ref{sec:discuss_jointfit}.
\begin{table}[h]
    \centering
    \begin{tabular}{cccccccc}
        Regression model  & $\widehat{\log(c_0)}$ & $\por$ & $\mu(\tau)$ & $\sigma(\tau)$ & $S$ & $\rmax$ & $\vpnm$ \Bstrut \Tstrut \\
        $v^{(1)}$   &  1.1095 & 4.4086 &   &   &   &   &\Bstrut \Tstrut  \\ \hline
        $v^{(2)}$   & -1.4714  & 0.71764 & -6.6531  & 0.9563  & -2.0561  &   &\Bstrut \Tstrut  \\ \hline
        $v^{(3)}$   & 6.7027  & &   &   &   &   & 1.7004 \Bstrut \Tstrut  \\ \hline
        $v^{(4)}$   &  8.8068 & -2.6707 &&   &   &   &  2.5019  \Bstrut \Tstrut  \\ \hline
        $v^{(5)}$   &  8.3394 & -2.8884 &  &  &   &  0.22851  & 2.5101\Bstrut \Tstrut  \\ \hline
        $v^{(6)}$   &  7.7435 & -1.5178 &  2.9228 & -0.17009  & -0.99981  &  & 2.7525 \Bstrut \Tstrut  
    \end{tabular}
    \caption{Coefficients of each descriptor
     determined for the regression models $v^{(1)},\ldots,v^{(6)}$
    for  the fit to the data of uncompressed paper.}
    \label{tab:uc_coefficients}
\end{table}
\begin{table}[h]
    \centering
    \begin{tabular}{cccccccc}
        Regression model  & $\widehat{\log(c_0)}$ & $\por$ & $\mu(\tau)$ & $\sigma(\tau)$ & $S$ & $\rmax$ & $\vpnm$ \Bstrut \Tstrut  \\ \hline
        $v^{(1)}$   &  2.225 & 5.2544 &   &   &   &  &  \Bstrut \Tstrut  \\ \hline
        $v^{(2)}$   & -8.3291  & 6.0792 & -0.97848  & 0.18181  & -6.7495  & &  \Bstrut \Tstrut  \\ \hline
        $v^{(3)}$   & 12.191  &  &   &   &   &  & 2.6676 \Bstrut \Tstrut  \\ \hline
        $v^{(4)}$   &  7.1945 & 3.7481 & &   &   & &  1.1189 \Bstrut \Tstrut  \\ \hline
        $v^{(5)}$   &  -0.59186 & 2.0003 &  &  &   & 2.0024 & 0.52999\Bstrut \Tstrut  \\ \hline
        $v^{(6)}$   &  -3.6529 & 5.0916 &  -0.62714 & 0.12062  & -5.7452  &  & 0.70949 \Bstrut \Tstrut  
    \end{tabular}
    \caption{Coefficients of each descriptor
     determined for the regression models $v^{(1)},\ldots,v^{(6)}$
    for  the fit to the data of compressed paper.}
    \label{tab:c_coefficients}
\end{table}
\begin{table}[h]
    \centering
    \begin{tabular}{cccccccc}
        Regression model  & $\widehat{\log(c_0)}$ & $\por$ & $\mu(\tau)$ & $\sigma(\tau)$ & $S$ & $\rmax$ & $\vpnm$ \Bstrut \Tstrut  \\ \hline
        $v^{(1)}$   &  0.75602 & 3.9832 & &   &   &   &   \Bstrut \Tstrut  \\ \hline
        $v^{(2)}$   & -1.4725  & 3.313 & -3.3552  & 0.54879  & -2.1554  &  &  \Bstrut \Tstrut  \\ \hline
        $v^{(3)}$   & 9.4227  &  &   &   &   &   & 2.1938 \Bstrut \Tstrut  \\ \hline
        $v^{(4)}$   &  5.9854 & 1.8556 &  &   &   &  &  1.2756 \Bstrut \Tstrut  \\ \hline
        $v^{(5)}$   &  2.9935 & 0.49562 & & &   &  1.302  &  1.2824 \Bstrut \Tstrut  \\ \hline
        $v^{(6)}$   &  4.8173 & 1.9791 & -0.93914  & 0.24517  &  -0.60351  &  & 1.1089 \Bstrut \Tstrut  
    \end{tabular}
    \caption{Coefficients of each descriptor
     determined for the models $v^{(1)},\ldots,v^{(6)}$,  
    fitted to the joint data set of uncompressed and compressed paper.}
    \label{tab:joint_fit_coefficients}
\end{table}

\subsection{Fitting of the regression model to the combined data set}
\begin{figure}[H]
    \centering
    \includegraphics[width=.95\textwidth]{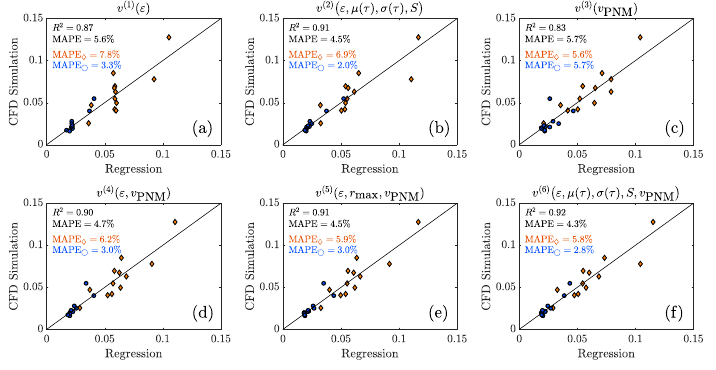}
    \caption{Scatter plots of volume flow rates obtained by CFD simulations vs the correspnding values predicted by the respective regression models based on the combined dataset. The coefficient of determination and the mean absolute percentage are displayed for both samples on each plot. Here, the regression models are fitted to the entire data set of uncompressed~(orange diamonds) and compressed~(blue circles) paper combined. Additionally, the mean absolute percentage errors evaluated on the subset of uncompressed and compressed data points are shown along with their corresponding symbol (diamond for uncompressed paper, circle for compressed paper).}
    \label{fig:regression_joint_fit_allmodels}
\end{figure}

Analogously to Figures~\ref{fig:FitDescriptors2} and~\ref{fig:regression_corrPNMs_fit3} in the main text, Figure~\ref{fig:regression_joint_fit_allmodels} shows scatter plots depicting the volume flow rates obtained by CFD simulations vs the corresponding values predicted by regression models based on the fit to the combined data of uncompressed and compressed paper.

\subsection{Details of the determination of the pore network}\label{sec:app-porenetwork}
The SNOW algorithm~\cite{gostick2017versatile} determines the positions of pore centers and pore volumes from 3D images as follows. 
For each 3D image, the Euclidean distance map~\cite{soille.2003} is acquired and the local maxima therein are recorded. 
This distance map is a gray-scale image in which every voxel of the pore space is assigned a value equal to the distance to the nearest voxel of the solid phase within the porous material. 
In order to eliminate superfluous local maxima in the distance map, we applied a Gaussian filter to the distance map prior to passing it to the SNOW algorithm.
The number of maxima registered by SNOW depends on  the radius ${R}$ of the spherical structuring element in the watershed segmentation, and the standard deviation $\sigma$ of the Gaussian filter applied to the distance map. 
To make the pore networks of the cutouts as comparable as possible, the values of $R = 3$ and $\sigma = 0.35$
were determined in an iterative process. Starting from the distance transform, we systematically tested different values of $\sigma$ in the range from 0 to 0.5, for $R=3$ and $R=4$, respectively, and multiple values for $R$ for $\sigma = 0.35$ to determine the values beyond which the amount of local maxima remained stable.
To find the boundaries between regions, the final set of local maxima are transformed into a set of markers. The marker-based watershed algorithm then determines distinct regions starting from the location of each marker~\cite{gostick2017versatile}. Each marker corresponds to exactly one region in the final segmented image. Subsequently, the centroid of each region serves as the coordinates of a pore.

The SNOW algorithm was applied for the combined internal and exterior pore space as described above, so that not only the internal pore space, but also the exterior pore space is split into regions through the segmentation procedure. 
All exterior regions that border the regions related to the internal pore space are assigned as inlet and outlet regions and will serve as inlets and outlet vertices in the transport simulation.
We ensured that none of the inlet or outlet vertices are directly connected to each other to prevent artificial pathways bypassing the inner pores in the network.

\subsection{Conductance through conduits}\label{sec:app-conduit}

Under the condition of slip-free incompressible laminar axis-symmetric flows, it is possible to provide analytical and semi-analytical expressions for the flow based on the Navier-Stokes equation.

\begin{figure}[h!]
	\begin{center}
  \includegraphics[width = .35\textwidth]{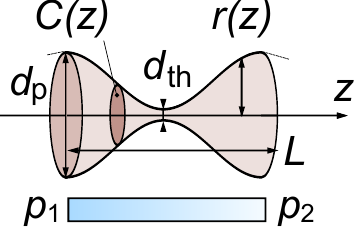}		
	\end{center}
	\caption{Axial-symmetric conduit shape with variable cross-sectional area along the flow direction $z$ from  left to right with a total pressure drop $\Delta p = p_2 - p_1$. Illustrated here is a conduit shape with additional mirror-symmetry in its converging-diverging shape.}\label{fig:flux_expressions}
\end{figure}

We assume that we are given an axial-symmetric conduit as in Figure~\ref{fig:flux_expressions}, which is placed within the positive quadrant of the coordinate system with one of its sides aligned with the $x$-$y$-plane, so that its extension along the $z$-axis is from $0$ to $L$.
For each $z\in [0,L]$, let $C(z) \subset \R^3$ be the circular conduit cross-section perpendicular to the flow axis, and $\mathcal{H}_2(C(z))$ the area of $C(z)$, given by means of the two-dimensional Hausdorff measure $\mathcal{H}_2$, see Figure~\ref{fig:flux_expressions}.
In steady state, the flux $Q = \partial V / \partial t$ through a rigid conduit with the area $\mathcal{H}_2(C(z))$ of the cross-section $C(z)$ is given by
\begin{equation}
	\label{eq:fl1}
   \frac{dp}{dz} = -16\pi^2 \eta  Q  \frac{I_p^*(z)}{\mathcal{H}_2(C(z))^2} + 2 \rho Q^2  \frac{1}{\mathcal{H}_2(C(z))^3}\frac{d\mathcal{H}_2(C(z))}{dz},
\end{equation}
where $\rho$ is the mass density and $\eta$ the dynamic viscosity of the flowing medium~\cite{akbari_viscous_2011}.
The area $\mathcal{H}_2(C(z))$ of the cross section $C(z)$ may vary along the conduit axis, see Figure~\ref{fig:flux_expressions}. 
The factor $I_p^*(z)$ is specific polar moment of inertia, defined as an integral across the cross-section $C(z)$:
\begin{equation}
	I_p^*(z) = \frac{1}{\mathcal{H}_2(C(z))} \int_{C(z)} \Vert x \Vert^2 \,\textup{d}\mathcal{H}_2(x),
\end{equation}
The two terms in Eq.~\eqref{eq:fl1} are referred to as friction term and inertia term, respectively.
The total pressure drop $\Delta p$ after the conduit at position $z=L/2$ is obtained by integrating Eq.~\eqref{eq:fl1} along the extension of the conduit along the flow axis:
\begin{eqnarray}
	\label{eq:conduitPressVsFlux}
	\Delta p  & = & -\int_{0}^{L} 16 \pi^2 \eta Q \frac{I_p^*}{\mathcal{H}_2(C(z))^2}\textup{d}z + \int_{0}^{L} 2\rho Q^2 \frac{1}{\mathcal{H}_2(C(z))^3}\frac{\partial \mathcal{H}_2(C(z))}{\partial z} \,\textup{d}z \\
	& = &  -\int_{0}^{L} 16 \pi^2 \eta Q\frac{I_p^*}{\mathcal{H}_2(C(z))^2}\textup{d}z + \rho Q^2 \left[ \frac{1}{\mathcal{H}_2(C(z))^2} \right]^{z=0}_{z=-L}
\end{eqnarray}
For conduits that exhibit a mirror-symmetry in addition to their axial-symmetry (as in Figure~\ref{fig:flux_expressions}) the second, \emph{i.e.}, the inertia term vanishes exactly.
Even without symmetry, the inertia term is typically much smaller compared to the friction term, so that only the friction term is considered further. With that, the flux $Q$ determines the pressure difference $p_1 - p_2$ across the conduit as
\begin{equation}
	\label{eq:conduitPressVsFluxFric}
	p_1 - p_2  = \int_{0}^{L} 16 \pi^2 \eta Q \frac{I_p^*}{\mathcal{H}_2(C(z))^2}\textup{d}z.
\end{equation}
Rearranging Eq.~\eqref{eq:conduitPressVsFluxFric}, the flux $Q$ can be conveniently expressed as 
\begin{equation}
	Q = g_{12} (p_1 - p_2)
\end{equation}	
with a so-called hydraulic conductance $g_{12}$ given by
\begin{equation}
	\frac{1}{g_{12}} =  16 \pi^2 \eta \int_{0}^{L} \frac{I_p^*}{\mathcal{H}_2(C(z))^2} \textup{d}z.
\end{equation}	

The factor $g_{12}$ can be considered as a shape factor that is purely related to the geometry of the conduit. The integration in $g_{12}$ has to be performed depending on the chosen shape of the conduit. Hence, shapes whose radial contour are defined by an analytical expression promise to yield a well defined conductance $g_{12}$. The above derivation is specifically for axially-symmetric conduit shapes. However, also for non-axially symmetric shapes, $g_{12}$ can be defined, see~\cite{akbari_viscous_2011} for more information.

\subsection{Impact of conduit shape on pore-to-pore flux}

\begin{figure}[h]
	\begin{center}
		\includegraphics[width = 0.5\textwidth]{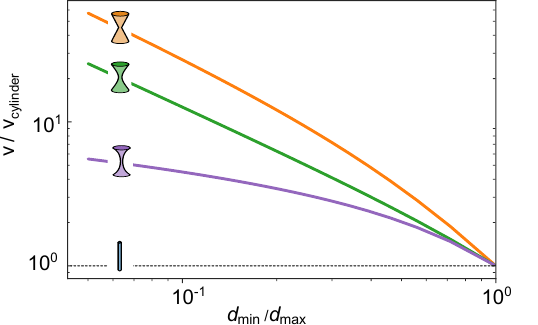}		
	\end{center}
	\caption{Dependence of the volume flow rates on the shape of the conduit between two connected pores. For all shapes, the same length is assumed. The basis converging-diverging shapes include cylinder (blue), hyperbolic cosine (purple), hyperbolic (green) and double cone (orange). For each basis shape, the ratio between the minimum diameter, $d_{\mathrm{min}}$ and the inlet diameter $d_{\mathrm{max}}$ is shown. Due to the symmetry of the assumed shapes, the inlet diameter equals the outlet diameter. The volume flow rates are given in relation to the volume flow rate through a cylinder of the same minimal diameter $d_{\mathrm{min}}$.  
    }\label{fig:flux_poreshape}
\end{figure}

The flow through a conduit strongly depends on the geometric shape of the conduit. 
Using the analytic formulas provided in~\cite{sochi_newtonian_2013}, we demonstrate in Figure~\ref{fig:flux_poreshape} that the flow rates increase the larger the discrepancy between the inlet diameter and the smallest diameter get.
The way in which the largest cross-section evolves into the smallest cross-section also influences the flow through the conduit. 
Going from a hyperbolic cosine shape (purple in Figure~\ref{fig:flux_poreshape}) to a linear shape corresponding to a truncated cone (orange in Figure~\ref{fig:flux_poreshape}) with length, minimum and maximum diameters held constant increases the flow rate by up to a factor of five.

\end{document}